\documentclass[conference]{IEEEtran}
\ifCLASSINFOpdf
  \usepackage[pdftex]{graphicx}
   \graphicspath{{Images/}}
\else
  \usepackage[dvips]{graphicx}
  \graphicspath{{Images/}}
\fi
\usepackage{subfig}

\usepackage{amsmath}
\usepackage{amssymb}
\usepackage{bm}

\newcommand{\bd}[1]{\mbox{\boldmath $#1$}}
\newcommand{\RR}[1]{\mathbb{R}^{#1}} 

\newcommand{\spectrum}[1]{\bm{\xi}_{#1}}
\newcommand{\scalarspectrum}[1]{\xi_{#1}}
\newcommand{\recspectrum}[1]{\bm{\hat \xi}_{#1}}

\DeclareMathOperator*{\argmax}{arg\,max}

\begin{document}
%
\title{A Spectral Model for Multimodal \\ Redshift Estimation}

\author{\IEEEauthorblockN{Sven D. K\"ugler, Nikolaos Gianniotis,  Kai L. Polsterer}
\IEEEauthorblockA{Astroinformatics Group \\ Heidelberg Institute for Theoretical Studies gGmbH\\
Schlo\ss-Wolfsbrunnenweg 35,
 69118 Heidelberg, Germany\\
Email: \{Dennis.Kuegler, Nikos.Gianniotis, Kai.Polsterer\}@h-its.org}
}


%


\maketitle


\begin{abstract}
We present a physically inspired model for the problem of redshift estimation.
Typically, redshift estimation has been treated as a regression problem
that takes as input magnitudes and maps them to a single target redshift.
In this work we acknowledge the fact that observed magnitudes may 
actually admit multiple plausible redshifts, i.e. the distribution of redshifts 
explaining the observed magnitudes (or colours) is multimodal. Hence, employing one of the standard regression models,
as is typically done, is insufficient for this kind of problem, as most models implement either one-to-one or many-to-one mappings. The observed multimodality of solutions is a direct consequence of (a) the variety of physical mechanisms that give rise to the observations, (b) the limited number of measurements available and (c) the presence of noise in photometric measurements. 
Our proposed solution consists in formulating a model from first principles
capable of generating spectra. The generated spectra
are integrated over filter curves to produce magnitudes which are then matched to the observed magnitudes.
The resulting model naturally expresses a multimodal posterior over possible redshifts,
includes measurement uncertainty (e.g. missing values) and
is shown to perform favourably on a real dataset. 
\end{abstract}


%
\IEEEpeerreviewmaketitle

\section{Introduction}

\nocite{*}
The exploration of the history of the universe has been mainly driven 
by the detection and investigation of highly-redshifted extragalactic sources, 
such as the quasi-stellar objects (QSO, \cite{1993ARA&A..31..473A}). The study 
of the distribution of these objects over space and time allows us to draw precise 
conclusions about how the universe was initially formed and developed since then
\cite{2010ASSP...14..255A}. Additonally, photometric redshifts have been used 
in the studies of galaxy clusters \cite{2008MNRAS.387..969A} and in constraining 
the galaxy luminosity function \cite{1996AJ....112..929S}.

Due to the extreme luminosity of quasars they are perfect traces for the early
 universe. Therefore, a significant time of research has been spent on
estimating their redshifts.
 While spectroscopic surveys are extremely precise in doing so, they are extremely 
time-intensive and cannot be used to study a large fraction of the objects known to date.
Instead, photometric surveys are used to infer knowledge about the nature and redshift of 
the quasars. Originally, this was done in a template-based way \cite{2000A&A...363..476B} 
and it was only rather recently that the number of data-driven approaches has increased drastically
(e.g.~\cite{2010MNRAS.406.1583W, 2011MNRAS.413.1395O, 2011MNRAS.418.2165L} and many more). In 
these works the main focus has been the comparison of methodologies.
A popular tool of the community is the random forest \cite{Breiman2001} due to its
reproducibility, precision and favourable computational complexity.

In this work, we acknowledge the fact that the redshift estimation is a regression problem
that admits multiple solutions, i.e. there can be more than just one redshift $z$ that explains
the observed magnitudes. If we look at the set of possible solutions $z$ as a distribution,
then we are acknowledging that this distribution can be multimodal (multiple distinct solutions)
as opposed to unimodal (single distinct solution). This manifestation of multiple distinct solutions
is the consequence of multiple factors:
\begin{itemize}
\item \textit{Different physical mechanisms}.  The aforementioned quasars 
are actually a preamble for a variety of observed phenomena that are thought to originate from 
similar sources, simply observed under different circumstances, e.g. viewing 
angles \cite{1995PASP..107..803U}.
The emission of light due to a loss of gravitational energy is common for all the 
sources. However, there are a lot of different effects that can contribute to the 
appearance of the spectrum, e.g. the central electron 
density or the black hole mass and spin e.g. \cite{2010ApJ...711...50T}.
In addition, it has been observed that the appearence of
the quasars changes with redshift \cite{2000MNRAS.311..576K}. 
The superposition of only these effects can lead to the observed 
multimodality and make a physical modeling of quasar 
emission very cumbersome.
\item \textit{Limited number of photometric measurements}. An abundant number of photometric measurements
could potentially help us identify a distinct redshift solution. However, with only a limited number measurements available (e.g. $11$ measurements do not suffice to pinpoint a unique redshift \cite{Wu2012}), ambiguities appear in the guise of multiple distinct solutions.
\item \textit{The presence of noise in photometric measurements may introduce multiple solutions}. 
Typically, the noisier the measurements, the more difficult it becomes
to pinpoint the correct redshift. However, the presence of noise does not only introduce uncertainty, but also distinct solutions. As an example, consider the case of two sets of
photometric measurements $\bd{g}_1$ and $\bd{g}_2$ which look very similar in all but one band in which they differ significantly.
Assume further that, due to this significant difference in one band, the two measurements  correspond to distinct redshifts $z_1$ and $z_2$.
If we were to increasingly add noise to the distinct band, we should observe that beyond a certain level of noise, solution $z_1$ starts becoming a likely candidate solution for measurement set $\bd{g}_2$; we should also notice that $z_2$ starts becoming a likely solution for $\bd{g}_1$.
In the extreme case, that the only distinct band was completely missing (or equivalently riddled by high uncertainty), the remaining observed bands would point out that there are two distinct solutions $z_1$ and $z_2$ for both $\bd{g}_1$ and $\bd{g}_2$.
\end{itemize}

The above factors have a tremendous impact on the uncertainty and the number of distinct redhsift solutions. However, in most works concerning redshift estimation, some of the standard regression models are employed
which are typically geared towards optimising an objective that consists in a sum-of-squares error or some variation thereof:
\begin{eqnarray}
E = \int \!\!\! \int \left( f(x) - t  \right)^2 \ p(t|x) \ p(x) \ dt \ dx \ ,
\end{eqnarray}
where $f(\cdot)$ is a regression model, $x$ are the data inputs and $t$ the data targets.
By rewriting this loss as the bias-variance decomposition \cite{Bishop1995},
we gain the following insight:
\begin{eqnarray}
E\!\!\!\!\! &=& \!\!\!\!\! \int \!\!\! \int \left( f(x) - t  + \mathbb{E}[t|x] - \mathbb{E}[t|x] \right)^2 \ p(t|x) \ p(x) \ dt \ dx \ \notag \\
& & + \left(t - \mathbb{E}[t|x] \right)^2  \ p(t|x) \ p(x) \ dt \ dx \ \notag\\
\!\!\!\!\!  &=& \!\!\!\!\! \int \!\!\! \int \left( f(x) - \mathbb{E}[t|x] \right)^2  + \left(t - \mathbb{E}[t|x] \right)^2 \ p(x)  \ dx \ , 
\label{eq:bias_variance}
\end{eqnarray}
where $\mathbb{E}[t|x]$ is the conditional expectation $\int t p(t|x) dt$ for a given $x$.
The first term  in Eq.~\eqref{eq:bias_variance} tells us that an optimal regression model $f(\cdot)$
is one that is as close as possible to the conditional average $\mathbb{E}[t|x]$.
The second term is the variance in $t$ due the presence of noise.
Hence, we can interpret the fitted model as a conditional Gaussian density $\mathcal{N}(f(x), \sigma^2)$
with the variance given by the second term \cite{MDN}.
Hence, standard regression models $f(\cdot)$ are not appropriate for the redshift estimation
problem where multiple redshift solutions are possible.

In the machine learning literature,  inverse problems with multiple solutions have been addressed
by two prevailing frameworks, namely the mixture density network architecture (MDN) \cite{MDN} and the hierarchical mixture of experts (HME) \cite{Jordan1994}.
In MDN the target variable is modelled with a mixture of Gaussians.
The parameters of the Gaussian mixture are parametrised by the outputs of the neural network.
Hence, the parameters of the Gaussian mixture are a function of the data inputs and this
results in a flexible model that adapts its distribution to the local characteristics
of the input space, e.g. by controlling the mixture weights we can locally adapt the distribution  modality. Similarly, HME also offers local adaptation by partitioning the data space
and allocating different experts (i.e. statistical learning model) to each partition.
This allows for building complex models out of simpler models that specialise on smaller regions
of the data space.

As aforementioned, in the case of redshift estimation, the multimodality of solutions is a consequence of the presence of noise and the limited number of measurements and thus not (necessarily) a characteristic of the data space, i.e.~\textit{multimodality is not a function of the data space}.
Instead, in this work, we formulate a simple model based on first principles.
The model incorporates simple physical considerations and states in a generative
fashion how observed magnitudes arise from spectra.
Observational noise is incorporated in a transparent way and the multimodality in the distribution
of redshift solutions arises naturally.

\section{Model formulation}

\subsection{Probabilistic PCA for uncertain spectra}

For the proposed approach, all spectra are preprocessed and are shifted to their
rest-frames (see Section \ref{sec:preprocessing}). Therefore, not all wavelengths are observed for all spectra.
This prevents us from extracting photometry by integrating the spectra over the filter curves
for any redshift.
In order to make this integration possible, we propose to  ``fill in" the unobserved wavelengths which we treat here as missing data.
To that purpose, we employ probabilistic principal component analysis (PPCA),
with a slight modification that allows us to deal with missing/unobserved data.
The idea behind using PPCA is to treat the observed, ``incomplete" high-dimensional spectra
as noise-corrupted versions of low-dimensional coordinates. In other words, by having at our disposal only ``incomplete"
spectra, we try to reconstruct a lower-dimensional space that explains the behaviour of the observed data. 
Once we have identified the lower-dimensional space, we can generate ``complete" spectra by mapping a low-dimensional coordinate into the high-dimensional space of spectra.

In the following we give a brief overview of how PPCA is formulated omitting details that can be found in the original formulation  in \cite{tipping1999probabilistic}.
Following PPCA, we take the view that the observed high-dimensional data $\spectrum{n} \in \RR{D}$ are the images of $Q$-dimensional $(Q<D)$ coordinates $\bd{\theta}_n \in \RR{Q}$ under a linear mapping plus additive Gaussian noise of covariance $\bd{S}_n$:
\begin{equation}
\spectrum{n} = \bd{W} \bd{\theta}_n + \bd{\mu} + \bd{S}_n \ ,
\end{equation}
where $\bd{W} \in \RR{D\times Q}$ and $\bd{\mu} \in \RR{D}$ define the linear mapping
from the low-dimensional space to the high dimensional data space,
and $\bd{S}_n\in \RR{D\times D}$ is a diagonal covariance matrix whose diagonal elements
are equal to the variance in the measurement of $\spectrum{n}$, i.e.~$\bd{S}_n= diag(\sigma_{n,1}^2, \dots, \sigma_{n,D}^2)$. As aforementioned,  we do not observe a spectrum $\spectrum{n}$ in its entire
wavelength range, i.e.~certain values $\scalarspectrum{ni}$ are unobserved. Our solution for dealing
with unobserved wavelengths is to set the unobserved $\scalarspectrum{ni}$ equal to a fixed value $\bar{\xi}_n$
and set the corresponding variance to a high value $\sigma_{n,i}^2 = \sigma_{high}^2$ , thus stating our ignorance for these unobserved values at wavelengths $i$.

Still following PPCA, the model is completed by imposing a Gaussian prior on the latent variables $\mathcal{N}(\bd{\theta}_n |\bd{0},\bd{I})$. This gives rise to the following log-likelihood:
\begin{equation}
\log \mathcal{L}(\bd{W},\bd{\mu}) = \log \prod_{n=1}^N \mathcal{N}(\spectrum{n} | \bd{W}\bd{\theta}_n + \bd{\mu}, \bd{S}_n)  \ \mathcal{N}(\bd{\theta}_n | \bd{0}, \bd{I}_D) \ .
\end{equation}
Treating the low-dimensional coordinates $\bd{\theta}_n$ as latent variables, PPCA formulates an expectation-maximisation algorithm. In the expectation step, one calculates the expected log-likelihood $\mathbb{E}_{q(\mathbf{\theta})}[ \log \mathcal{L} ]$ where the expectation is taken
over the posterior of the latent variables $q(\bd{\theta}_n)$ which due to the linear-Gaussian structure of the model is a Gaussian density, $q(\bd{\theta}_n) = \mathcal{N}(\bd{\theta}_n | \bd{m}_n,\bd{C}_n)$.
Hence in the expectation step we need the posterior mean and covariance  of each latent variable. These quantities are calculated as:
\begin{align}
\bd{C}_n &= \left( \bd{I} + \bd{W}^T \bd{S_n}^{-1} \bd{W} \right)^{-1} \\
\bd{m}_n &= \bd{C}_n \bd{W}^T \bd{S}_n^{-1} (\spectrum{n}-\bd{\mu})
\end{align}
and can be contrasted to the corresponding equations $(25)$ and $(26)$ of the standard PPCA found in \cite{tipping1999probabilistic}. Armed with these posteriors, we are in position to calculate $\mathbb{E}_{q(\mathbf{\theta})}[ \log \mathcal{L}(\bd{W},\bd{\mu})]$. In the maximisation step, we optimise\footnote{Just like in the original PPCA, \bd{\mu} is set equal to the sample mean of the data.} $\bd{W}$ by  calculating the gradient $\frac{\partial }{\partial \mathbf{W}}\mathbb{E}_{q(\mathbf{\theta})}[ \log \mathcal{L}(\bd{W},\bd{\mu})  ]$
and employing it in a gradient-based optimiser.

\begin{figure}[!t]
\centering

\subfloat{%
\includegraphics[width=0.47\textwidth]{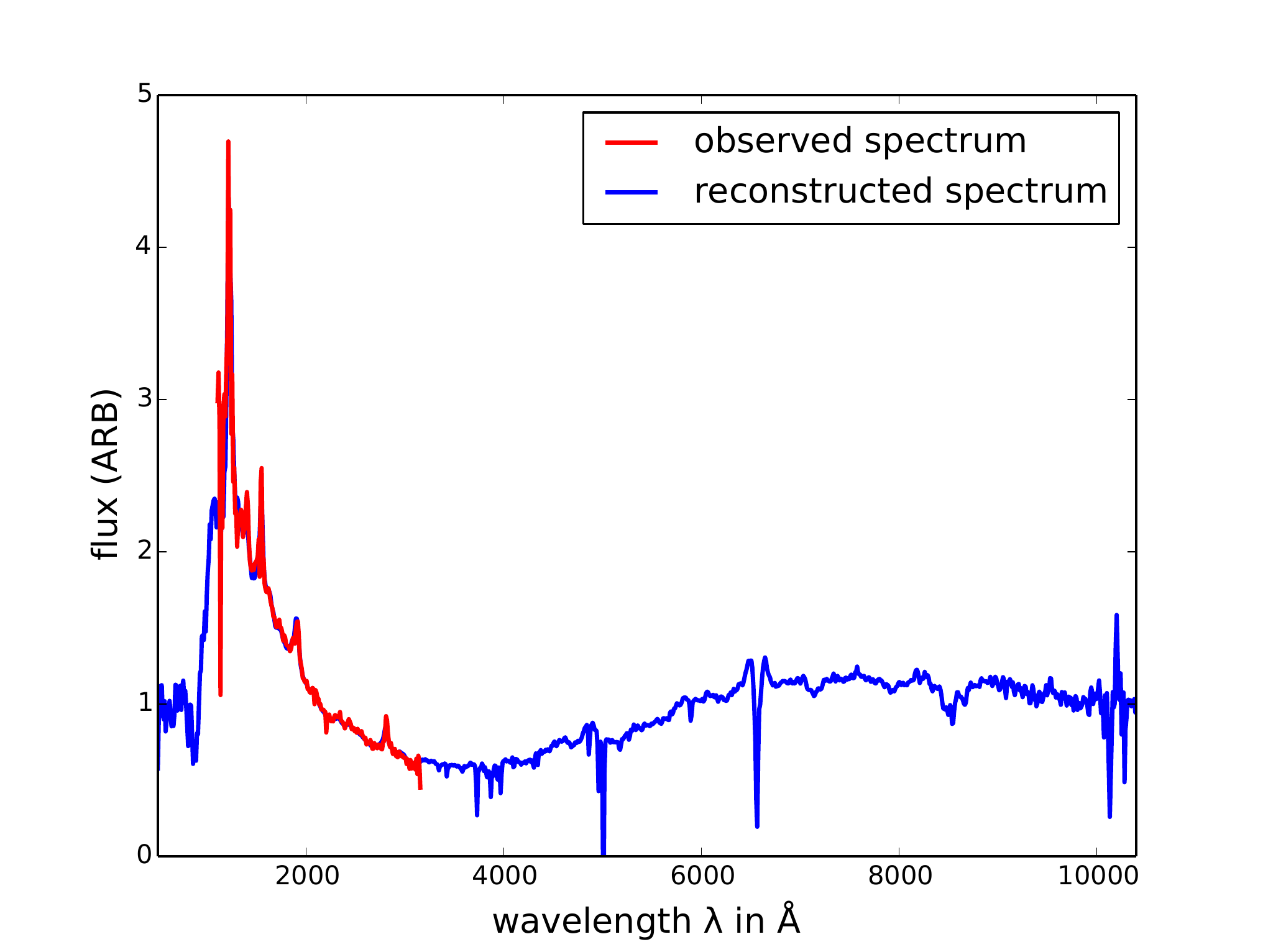}
\label{fig:ppca_example}
}
\\
\subfloat{%
\includegraphics[width=0.47\textwidth]{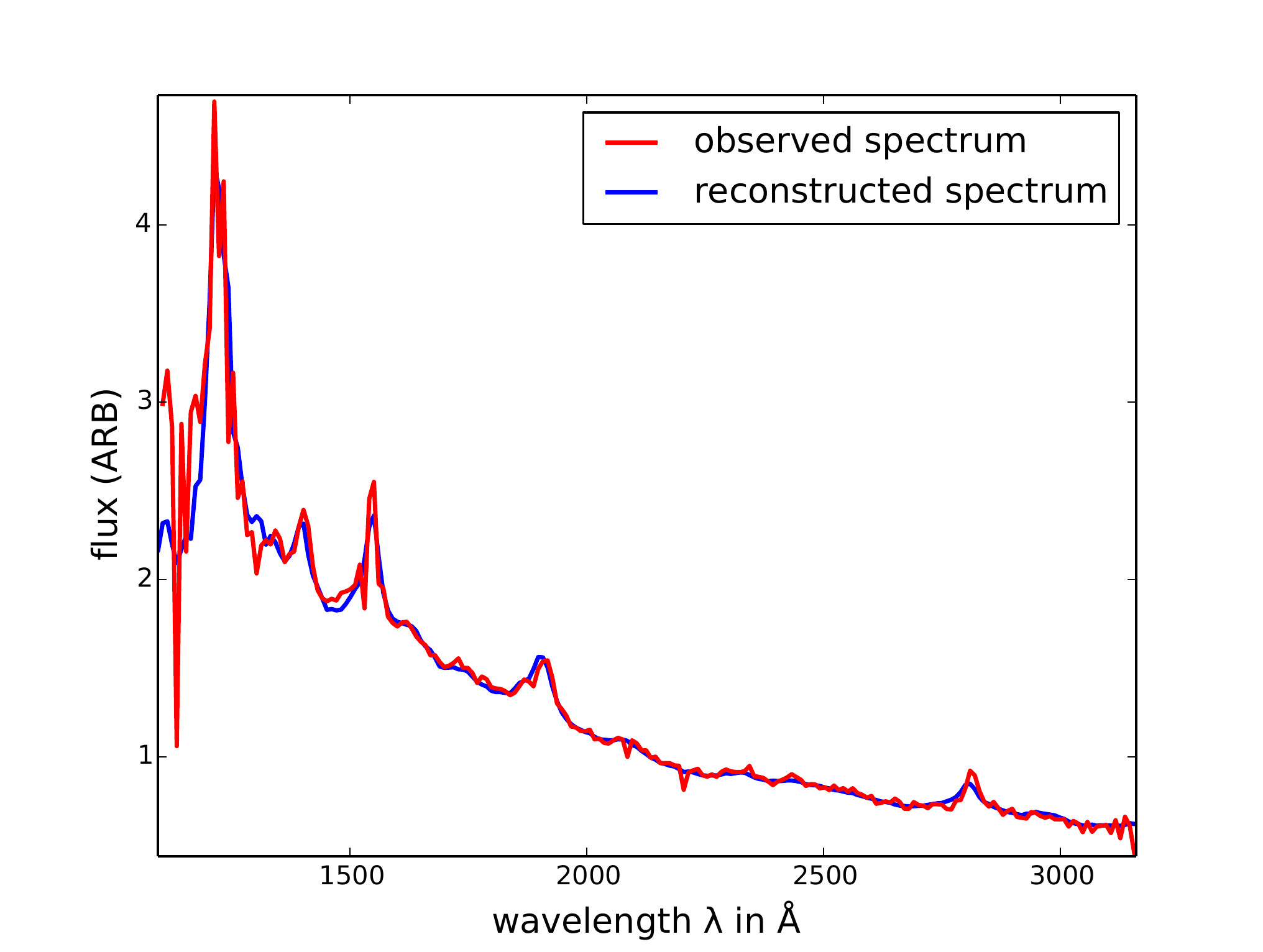}
\label{fig:ppca_example_zoomed}
}
\caption{Above: observed ``incomplete" spectrum $\spectrum{}$ plotted in red, reconstructed and ``complete'' spectrum $\recspectrum{}$ obtained by PPCA plotted in blue. Below: same as before, but we zoom in the region where observed and reconstructed spectrum overlap. \label{fig:ppca}}
\end{figure}

Once the expectation-maximisation algorithm has converged, we are able to map previously unseen (out-of-sample) ``incomplete" spectra $\spectrum{\star}$, with covariance matrix $\bd{S}_{\star}$, to the low-dimensional space by:
\begin{equation}
\bd{\theta}_{\star} = \left(\bd{I} + \bd{W}^T \bd{S}_{\star}^{-1} \bd{W}\right)^{-1} \bd{W}^T \bd{S}_{\star}^{-1} (\spectrum{\star}-\bd{\mu}) \ .
\end{equation}
A reconstruction for $\spectrum{\star}$ can be obtained by mapping back to the data space:
\begin{equation}
\recspectrum{\star} = \bd{W}\bd{\theta}_{\star} + \bd{\mu} \ .
\end{equation}
Figure \ref{fig:ppca} illustrates how an observed, out-of-sample, spectrum $\spectrum{\star}$ is reconstructed as a ``complete" spectrum $\recspectrum{\star}$.

\subsection{Physical model}

In the previous section, we briefly described how PPCA embeds noise-corrupted spectra
in a low-dimensional space and how ``complete" reconstructions of spectra can be generated. 
In fact, we regard the PPCA as a generative model, parametrised by 
the low-dimensional coordinate $\bd{\theta}\in\RR{Q}$, that generates synthetic spectra
$\recspectrum{}(\bd{\theta})$ that closely resemble observed spectra $\spectrum{}$.
In the following, we detail how this generative model can be exploited
in a redshift regression model.

Observed photometric magnitudes are produced by a spectrum $\recspectrum{}(\bd{\theta})$, generated by our PPCA model, of an unknown $\bd{z}$ which we want to estimate. 
Furthermore, photometric magnitudes are modelled as the integration of the spectrum
over filter curves. Thus, the flux $\mathcal{I}$ in a band $b$ is computed as:
$$\mathcal{I}_b(\theta,z) = \frac{\int_0^\infty \lambda \recspectrum{}(\bd{\theta})(\lambda/(z+1))
f_b(\lambda)d\lambda}{\int_0^\infty \lambda 
f_b(\lambda)d\lambda}.$$
Since the spectra are discrete the transformation \mbox{$ \recspectrum{}(\bd{\theta}) (\lambda/(z+1))$} cannot
be continuously done. This is why, we use
the replacement
$$\tilde{\lambda} = \lambda/(z+1)$$
$$\frac{d\tilde{\lambda}}{d\lambda} = 1/(z+1) \Rightarrow
d{\lambda}= (z+1)d\tilde{\lambda}$$
and thus
\begin{equation}
\begin{split}
\mathcal{I}_b(\bd{\theta},z) & = \frac{\int_0^\infty (z+1)\tilde{\lambda}
\recspectrum{}(\bd{\theta})
f_b((z+1)\tilde{\lambda}) (z+1)d\tilde{\lambda}}{\int_0^\infty
(z+1)\tilde{\lambda} f_b((z+1)\tilde{\lambda})(z+1)d\tilde{\lambda}} \\
& \stackrel{\tilde{\lambda} \rightarrow \lambda}{=}
\frac{\int_0^\infty \lambda
\recspectrum{}(\bd{\theta}) f_b((z+1)\lambda)d\lambda}{\int_0^\infty
\lambda f_b((z+1)\lambda)d\lambda}.\\
\end{split}
\end{equation}
Effectively, we have now pushed the redshift from the discontinuous spectrum to
the continuous filter bands by replacing $\tilde{\lambda} \rightarrow
\lambda$. However, these can be easily approximated with an analytical function,
here with a linear regression model with basis functions 
\begin{equation}
f(\lambda)=\sum_{c=1}^{N_{comp}} v_c\exp
\left(-0.5\left(\frac{\mu_c-\lambda}{\sigma_c}\right)^2\right) \ , 
\end{equation}
with $v_c$, $\mu_c$, $\sigma_c$ being the weights, the means and the widths of each of
the $N_{comp}$ RBF basis functions. 

In order to compute now the expected flux from our model, we 
approximate this integral as a regular Riemann sum, where the bin width
$\Delta$ is given by the distance between two regularly sampled grid points, as
described in the preprocessing (see Section \ref{sec:preprocessing}). Finally, the flux in band $b$ is
computed as 
\begin{equation}
\begin{split}
\mathcal{I}_b(\theta,z) & \approx \frac{\Delta\sum^D_d \lambda_d
\recspectrum{}(\bd{\theta}) f_b((z+1)\lambda_d)}{\Delta\sum^D_d
\lambda_d f_b((z+1)\lambda_d)} \\
& = \frac{\sum^D_d
\lambda_d \recspectrum{}(\bd{\theta}) f_b((z+1)\lambda_d)}{\sum^D_d
\lambda_d f_b((z+1)\lambda_d)}.\\
\end{split}
\end{equation}
In summary, we know how the flux in a band $b$ for a spectrum generated by PPCA
coordinates $\bd{\theta}$ and redshift $z$ can be computed. Now, all we have to do is to
convert the observed magnitudes to equivalent fluxes in the
spectra\footnote{Note that we prefer to work in flux space, as the PCA might
well return also negative spectra, which are non-physical, but can still occur as
part of the optimization process.}, $10^{-0.4(T_b-ZP_b)}$,
where $M_b$ denotes the magnitude and $ZP_b$ is the zero-point\footnote{In our numerical experiments
zero-points are arbitrary. In general, for other data, we need to gauge them correctly.} for band $b$.
Lastly, we need to multiple the flux with an arbitrary scaling constant $s$, 
in order to accommodate for the difference in average flux, i.e. $s\mathcal{I}_b$.

The free model parameters are $s$, $\bd{\theta}$ and $z$.
Though in principle coordinate vector $\bd{\theta}$ is  continuous,
we found out in preliminary  numerical experiments that it was
very easy to overfit it. The reason of overfitting is because,
without imposing any control on it, coordinate $\bd{\theta}$ can move away
from the region in $\RR{Q}$ occupied by the PPCA-projections of the spectra. This has
as a consequence that $\bd{\theta}$ is mapped to arbitrary 
(physically implausible) spectra in its attempt to explain the observed fluxes
and hence $\bd{\theta}$ is overfitted. In order to circumvent this problem\footnote{An alternative solution would be to impose a penalty term on the continuous
parameters $\bd{\theta}$ that penalises distance from the data populated region in $\RR{Q}$.},
we choose to model parameter  $\bd{\theta}$ as a discrete parameter
that takes values in the set $\{\bd{\theta}_1, \dots, \bd{\theta}_{N} \}$,
the low-dimensional projections of the observed spectra $\spectrum{1}, \dots, \spectrum{N}$.
The low-dimensional projections $\bd{\theta}_n$ can be interpreted
as a discrete set of low-dimensional coordinates that give rise
to a set of spectra $\recspectrum{n}$.

Similarly, we also noticed
that gradient optimisation of $z$ is not practical. The objective function (likelihood in the next section)
is plagued by multiple local optima hence gradient optimisation gets easily trapped.
Hence, we also optimise $z$ by searching on a regular grid, see Table \ref{tab:params}.

\subsection{Model likelihood}

Assuming Gaussian noise on the observed data,
we define the following likelihood function for our model:
\begin{equation}
p(\bd{M}| \bd{\theta}, z, s) = \prod_{b} \mathcal{N}\left(10^{-0.4(M_b-ZP_b)}| s\mathcal{I}_b\left(\bd{\theta} ,z\right),{\sigma_b}^2 \right) \ ,
\end{equation}
where $\bd{M}$ is the vector of observed magnitudes, and $\sigma_b^2$ is respective variance in band $b$.


We complete the model by defining priors on the free parameters.
For scaling\footnote{Note that scaling can be omitted by 
optimizing colors instead of bands, then of course the input dimension would decrease 
by one accordingly.}, we impose the uninformative prior:
\begin{equation}
p(s) = \frac{1}{\log s_{max}-\log s_{min}}\frac{1}{s} \ ,
\end{equation}
while for the redshift and coordinates a uniform prior is assumed respectively.
Given observed magnitude data $\bd{M}$, we compute the posterior of $z$ by integrating over the 
discrete set of coordinates $\bd{\theta}_n$ and normalising:
\begin{equation}
\begin{split}
p( z_i | \bd{M}) &= \frac{\int \sum_{n} 
p(\bd{M}| \bd{\theta}_n, z_i, s)  p(\bd{\theta}_n) p(z_i) p(s) ds}
{\sum_{z_j} \int  \sum_{n} 
p(\bd{M}| \bd{\theta}_n, z_j, s)  p(\bd{\theta}_n) p(z_j) p(s) ds} \\
& \approx
\frac{\sum_{s_j} \sum_{n} 
p(\bd{M}| \bd{\theta}_n, z_i, s_j) p(\bd{\theta}_n) p(z_i) p(s_j) }
{\sum_{z_j} \sum_{s_j}  \sum_{n} 
p(\bd{M}| \bd{\theta}_n, z_j, s_j) p(\bd{\theta}_n) p(z_j) p(s_j) } \ , \\
\end{split}
\label{eq:posterior}
\end{equation}
where the integration over the scaling parameter $s$ is approximated by a sum
on a regular grid. We summarise  the evaluation grid of the discretised model parameters in \mbox{Tab.\,\ref{tab:params}}.

\begin{figure}
\centering
\includegraphics[width=.5\textwidth]{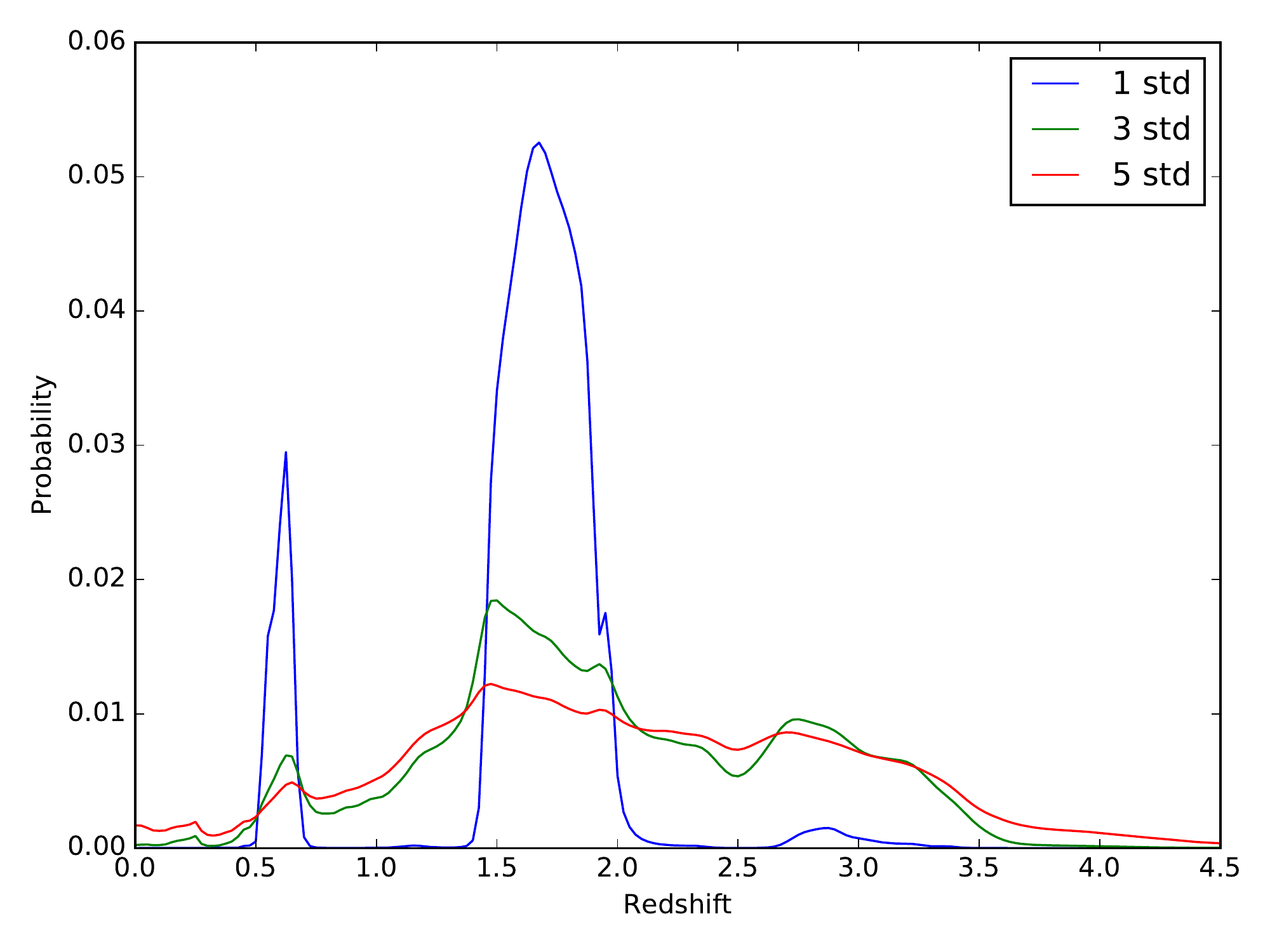}
\caption{The figure illustrates the multimodal posterior distribution calculated by the proposed model, and how this posterior changes in the presence of increasing noise $\sigma_b$ in the observed magnitudes.
The plotting range of the redshift is limited for illustration purposes.
\label{fig:increasing_noise}}
\end{figure}

We note that the posterior in Eq.~\eqref{eq:posterior} is a probability
mass function (PMF) defined on a discrete support, while the target redshifts $z$ in the
dataset take their values in a continuous interval. Hence, we cannot evaluate the posterior in  Eq.~\eqref{eq:posterior}  on arbitrary values. We therefore convert the 
PMF posterior into a piecewise uniform distribution.
That is, we define a uniform distribution for each interval between two grid points
and scale it by $p( z_i | \bd{M})$.
The new probability density distribution of our model simply reads:
\begin{equation}
 p_{spec}(z|\bd{M}) = \frac{1}{0.025  } \frac{p( z_i | \bd{M})}{\sum_{z_j} p( z_j | \bd{M})} \ , 
\label{eq:posterior_pc}
\end{equation}
for $z \in [0.025(i-1),0.025i]$, i.e. $z$ belongs to the $i-$th interval of the grid in Table \ref{tab:params}. The normalisation constant results from the fact that the distribution $p_{spec}$
consists of uniform distributions all of interval length $0.025$;
each uniform distribution is scaled by $p( z_i | \bd{M})$ and thus we also need to normalise by the sum 
$\sum_{z_j} p( z_j | \bd{M})$. 

Finally, if a point estimate is required from the proposed model, 
we use to that end the highest mode $\argmax_z p_{spec}(z)$.

\begin{table}[!t]
\caption{Evaluation parameters \label{tab:params}}
\centering
\begin{tabular}{lll}
Parameter & Regular grid & Grid points \tabularnewline
\hline
scaling $s$ & $\{0.5,0.525,\dots,2.0\}$ & 60 \tabularnewline
coordinate  & $\{ \bd{\theta}_1, \dots, \bd{\theta}_N \}$ & $5000$ \tabularnewline
redshift $z$ & $\{0.0,0.025,\dots,5.5\}$ & 220 \tabularnewline
\end{tabular}
\end{table}

%

\subsection{Demonstration of model behaviour} 

The posterior calculated in Eq.~\eqref{eq:posterior_pc} has the following benefits: first,
the model expresses a multimodal distribution over the possible redshifts $z$ that explain the observed magnitudes $\bd{M}$.
Secondly, the posterior changes in response to the presence of noise in the observations
$\bd{M}$.
This is illustrated in Fig.~\ref{fig:increasing_noise} on a data item from the dataset described in Section \ref{sec:data}.
Finally, we have control over the prior on $z$. In this work, we choose a uniform prior.

\begin{figure}
\centering
\includegraphics[width=.53\textwidth]{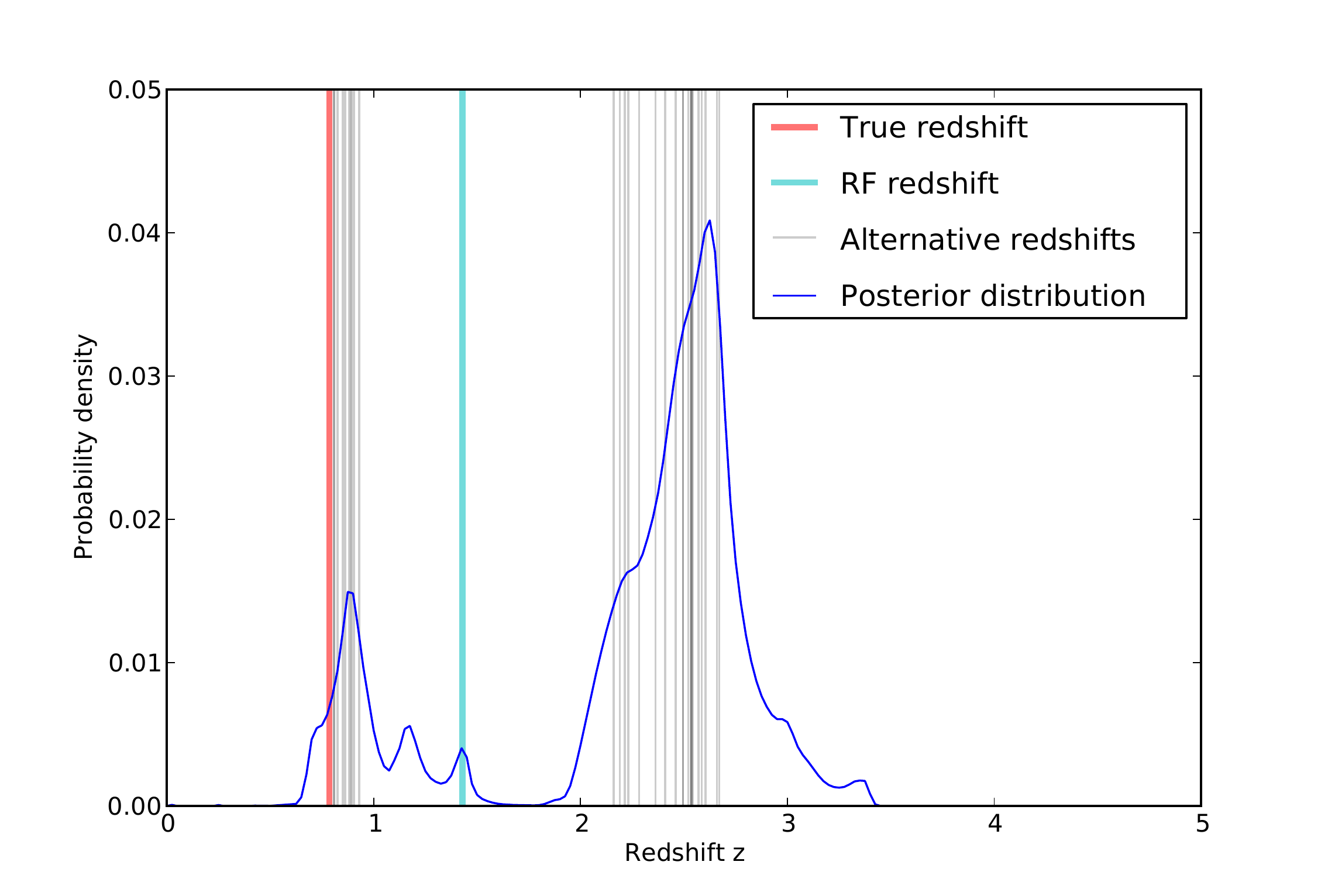}
\caption{For a given data item the actual (spectroscopic) redshift is shown in red. The redshift obtained by random forest regression  is shown in cyan. In addition, alternative redshifts, of objects that are very similar in terms of colour, are shown as grey lines. In blue we plot the density of the proposed model. \label{fig:example}}
\end{figure}

In Fig.~\ref{fig:example}, we pick a test object (from the dataset described in Section \ref{sec:data}) whose colour admits
multiple redshifts as possible solutions. 
Though these multiple alternative redshifts are not directly available, we
make the assumption  that they can be approximately recovered
as the redshifts that belong to objects that are closest (in Euclidean sense)
to the test object in terms of colour. These retrieved alternative redshifts
are plotted as grey lines. Hence, the first thing to note is that the grey lines
are clustered around two locations. This tells us that that objects with very similar colours may correspond to really
different redshifts, i.e.~this is evidence that in actual fact a colour can be associated with multiple distinct redshifts.
Amongst these redshifts, we also plot the true redshift as a red line,
which is known in this case as spectral data are available for the test object.
The posterior distribution $p_{spec}(z)$ obtained from our model, given in Eq.~\ref{eq:posterior_pc},
is plotted as a blue line. It is very pleasing to see that both dominating modes of the model posterior 
overlap with the alternative redshifts (grey lines).
The plot clearly shows that both dominating modes of the model posterior are justified:
the left-most dominating mode is close to the true redshift (red line), while the right-most dominating
mode explains the presence of an alternative distinct group of redshifts (grey lines).
We also plot the prediction of the random forest (trained on data described in Section \ref{sec:data})
as a line in cyan. As previously explained, the random forest cannot cope with multimodality
and hence its prediction is a compromise of the multiple modes. 
In this particular case, it leads to a prediction located in a region where the probability density is low, i.e.
in a region where no grey lines are present.

\section{Numerical experiments}

\subsection{Data} \label{sec:data}

\begin{figure}
\centering
\includegraphics[width=.5\textwidth]{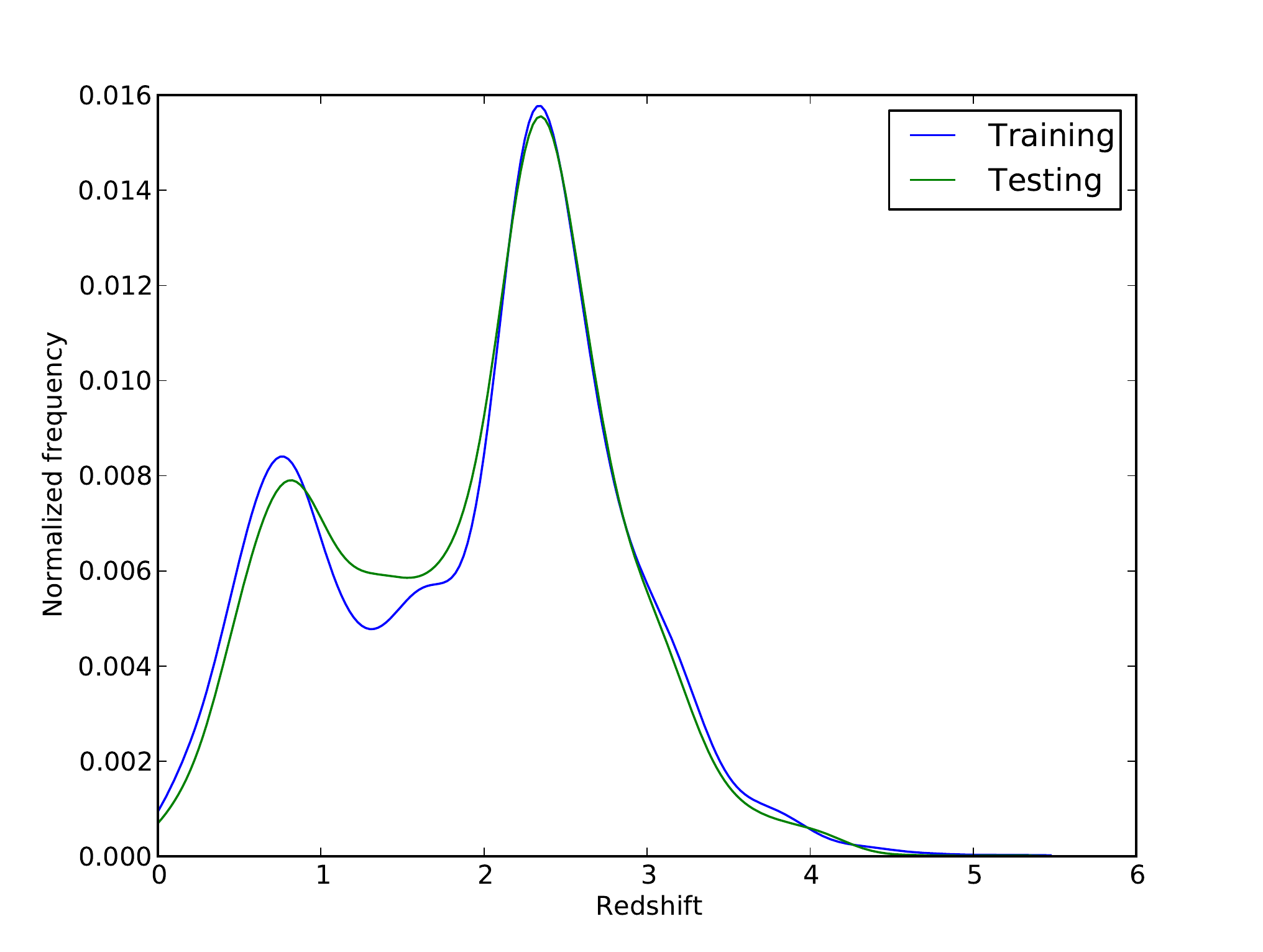}
\caption{Redshift distribution of training and test data.\label{fig:redDist}}
\end{figure}

We demonstrate
our proposed approach on a small subset of quasars contained in the BOSS catalogue. First, we 
extract 7506 randomly selected quasar spectra from BOSS which we divide into a training (5000)
and test set (2506). The redshift distributions are shown in Fig.~\ref{fig:redDist}. The 
idea is now to extract the photometry directly from the spectra instead of using their 
observed direct photometric counterparts. This way of approaching the problem has the following 
advantages:
\begin{itemize}
\item no calibration of the zero points needed
\item no uncertainties in the observables (spectra are considered noiseless)
\item full control over how data have been generated.
\end{itemize}
One of the downsides of using the spectra is that only part of the $u$ band is covered
and thus we have only $3$ colors at our disposal for inferring redshift (in the presented case these will be 
the three independent colors $g-r$, $g-i$, $g-z$). Note that in our methodology the fluxes
themselves are used instead of colours and therefore the data presented to our algorithm are
four dimensional. This is not an advantage as our model contains an additional scaling that
has to be optimized. 

The data are randomly split into a training and testing set of
$N=5000$ and $N_{test}=2506$ objects respectively.
In our approach, the training set is used to train the PPCA
and project the $5000$ spectra $\spectrum{n}$
to the the low-dimensional space $\RR{Q}$ and
in order obtain low-dimensional coordinates $\bd{\theta}_n$.

\subsection{Preprocessing} \label{sec:preprocessing}

All the required spectra are downloaded from the SDSS 
server. 
In a first step,
all  spectra are binned with a binning factor of 10 according to the following rules:
$$\lambda^j_{new} = \frac{1}{10}\sum_{i=10j}^{10(j+1)}\lambda^i_{old}$$
$$f^j_{new} = \frac{\sum_{i=10j}^{10(j+1)}f^i_{old} (\Delta f^i_{old})^{-2}}{\sum_{i=10j}^{10(j+1)}(\Delta f^i_{old})^{-2}}$$
$$\Delta f^j_{new} = \frac{1}{\sum_{i=10j}^{10(j+1)}(\Delta f^i_{old})^{-2}}$$
where $\lambda$, $f$ and $\Delta f$ are the wavelength, the spectral flux and the 
error of the spectral flux respectively. Subsequently, the spectra are shifted into 
their rest-frame and the flux values are extracted on a fixed grid ($\lambda \in [500,10400]$ in 1000
equally spaced steps) using spline interpolation. As aforementioned,
missing parts are given a value of  $\bar{\xi}_n$ and a standard deviation of $\sigma_{high}$.

\subsection{Performance Criteria} \label{sec:criteria}

We compare the algorithms using the following criteria:
\begin{itemize}
\item Root mean squared error, $RMSE$.
A way of measuring error in redshift regression problems is 
the normalized redshift deviation
$\Delta z_{norm} = \frac{z_{reg}-z_{true}}{1+z_{true}} $.
We employ the root mean square error ($RMSE$) 
as a performance criterion calculated as:
\begin{equation}
RMSE= \sqrt{ \frac{1}{N_{test}} \sum_{n=1}^{N_{test}} \Delta z_{norm,n}^2 } \ .
\end{equation}
\item Median absolute deviation, $MAD$:  this frequently used measure reads:
\begin{equation}
\frac{1}{N_{test}}\sum_{n=1}^{N_{test}} |\Delta z_{norm,n} - median(\Delta z_{norm})| \ ,
\end{equation}
and is less susceptible to highly deviating objects.
\item Likelihood, $\frac{1}{N}\mathrm{logL}_{\mathrm{True}}$.
This measure expresses how well the data are explained under a model, i.e. the model
learnt by each of the three candidate algorithms (see Section \ref{sec:algorithms}). We  compute under each model
the likelihood averaged over all data items in the test set:
\begin{equation}
\frac{1}{N}\mathrm{logL}_{\mathrm{True}} = \frac{1}{N_{test}} \sum_{n=1}^{N_{test}}p_{model} \left(z_{n;\,true}\right) \ ,
\label{eq:logL}
\end{equation}
where $p_{model} \in \{p_{GP},p_{RF},p_{spec} \}$.

\end{itemize}

\subsection{Algorithms} \label{sec:algorithms}

The following algorithms are put to the test:

\begin{itemize}
\item Proposed approach. We set the number of embedding dimensions for PPCA to $Q=10$.
Concerning the treatment of missing values in the observed spectra, we
set $\bar{\xi}_n$ equal to the average of observed values of spectrum $\spectrum{n}$,
and use a fixed $\sigma_{high}=1000$ for all spectra.

The $RMSE$ and $MAD$ performance criteria described in Section \ref{sec:criteria},
require that models deliver a point prediction. For the proposed model,
we take its point prediction to be the highest mode of the posterior $\argmax_z p_{spec}(z)$.

\item Random forest. Due to its popularity and success in the astronomical community,
we include the random forest (RF) \cite{Breiman2001} as a candidate algorithm. The number of trees is set to $1000$.
For comparison purposes, we derive a likelihood function for the RF.
The likelihood is simply defined as a Gaussian distribution with a mean given
by the prediction $z_{RF}$ of the RF and the standard deviation calculated by the residuals 
(here $\sigma=0.30$).
Thus we obtain, $p_{RF}(z)=\mathcal{N}\left(z|z_{RF},\sigma^2\right)$. 
Alternatively, we also optimise $\sigma$ so that criterion $\frac{1}{N}$logL$_{\mathrm{True}}$ is
optimized ($\sigma=0.76$), please see Section \ref{sec:criteria}.
%
\item Gaussian Process. We include Gaussian process (GPs) \cite{Rasmussen} in our comparison since
it is a flexible model that enjoys automatic regularisation and outputs a Gaussian predictive density
$p_{GP}(z)$. We employ the standard RBF kernel.
\end{itemize}

\subsection{Results}
\label{sec:results}

In Table \ref{tab:results}, we show how the three algorithms fare according to the performance criteria.
For the commonly used $RMSE$, we can see that the GP and RF perform very similar and significantly better than the proposed approach. However, in the other two criteria the presented algorithm performs 
much better than its competitors.  
We report in detail the results for each criterion.

\begin{figure}
\centering
\includegraphics[width=.5\textwidth]{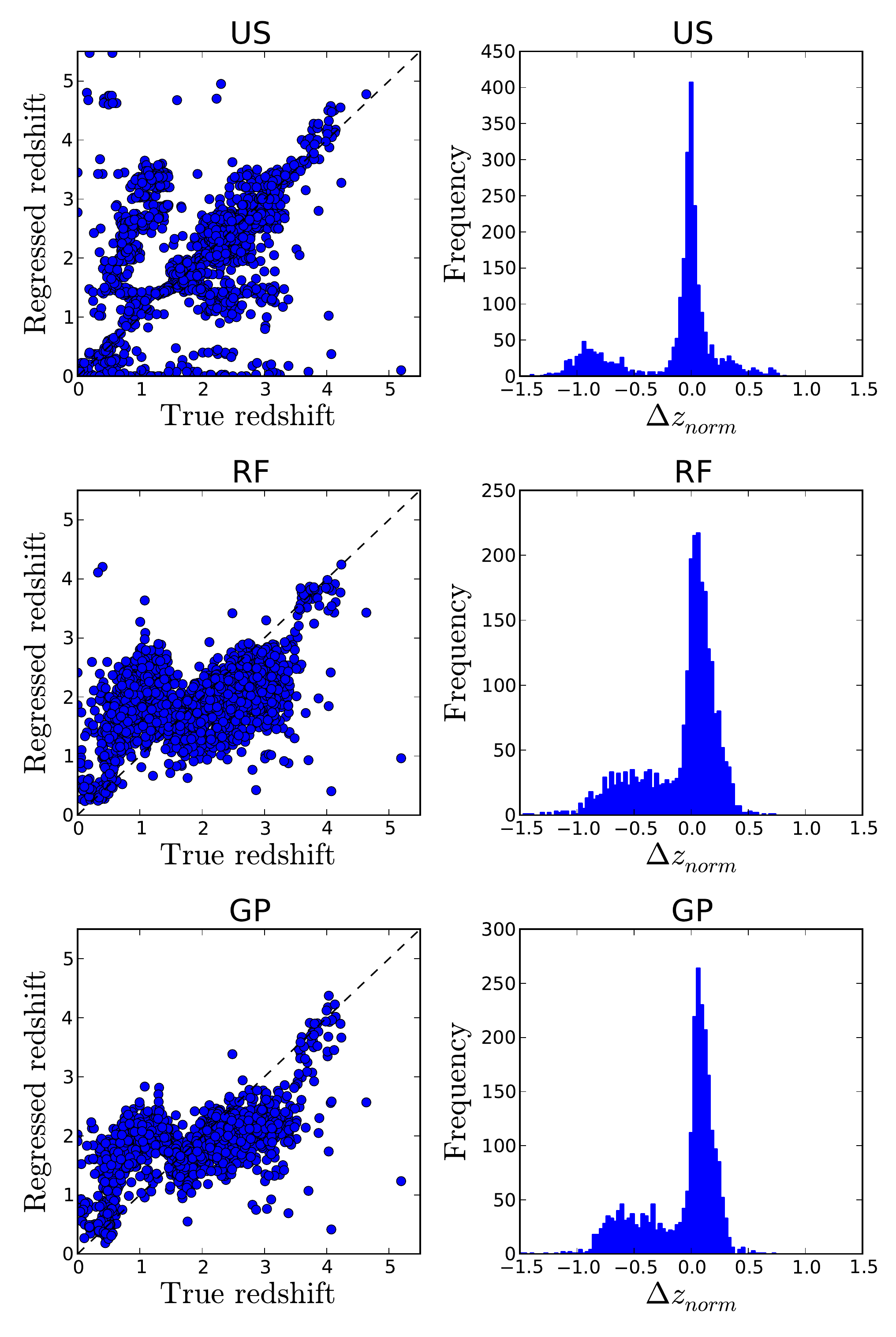}
\caption{Each row shows the results of the experiment noted on top of each plot (our model, random forest and Gaussian process). On the left side the value obtained by the regression algorithm is plotted against the actual value. On the right one can see a histogram of $\Delta z_{norm} = \frac{z_{reg}-z_{true}}{1+z_{true}}$. One can clearly see that for the proposed model the peak is much sharper, but the left wing is much more pronounced than for the other two algorithms.\label{fig:allResults}}
\end{figure}

In Fig.\,\ref{fig:allResults}, the regressed redshift is plotted against
the true one and additionally a histogram over $\Delta z_{norm}$ is shown for each of the algorithms.
On a first look, we  clearly see why the $RMSE$ is much worse for the proposed model.
While for the GP and RF the points are closer to the diagonal line (left column of plots in Fig.~\ref{fig:allResults}), there are some very drastic
deviations apparent in our algorithm. This behaviour can be explained by the fact that 
the GP and RF adapt to the distribution of $z$ in the dataset, i.e. 
they learn to a certain degree that most redshifts fall in the interval.
This is reasonable, if the redshift distribution is \emph{similar} for training and testing.
If this cannot be guaranteed (as in most  realistic settings, since the observational biases between surveys can be different), this will effectively lead to an amplification of this bias and thus even worse predictions will be produced, as shown later on.
Our model adopts as a prior the uniform distribution over redshifts, but of course we can influence this behaviour by choosing as a prior the distributions of redshifts in the dataset.
If we do so, the predictions get closer to the diagonal (not shown in figures),
and we obtain an $RMSE$ of $0.400$.

\begin{figure}
\centering
\includegraphics[width=.41\textwidth]{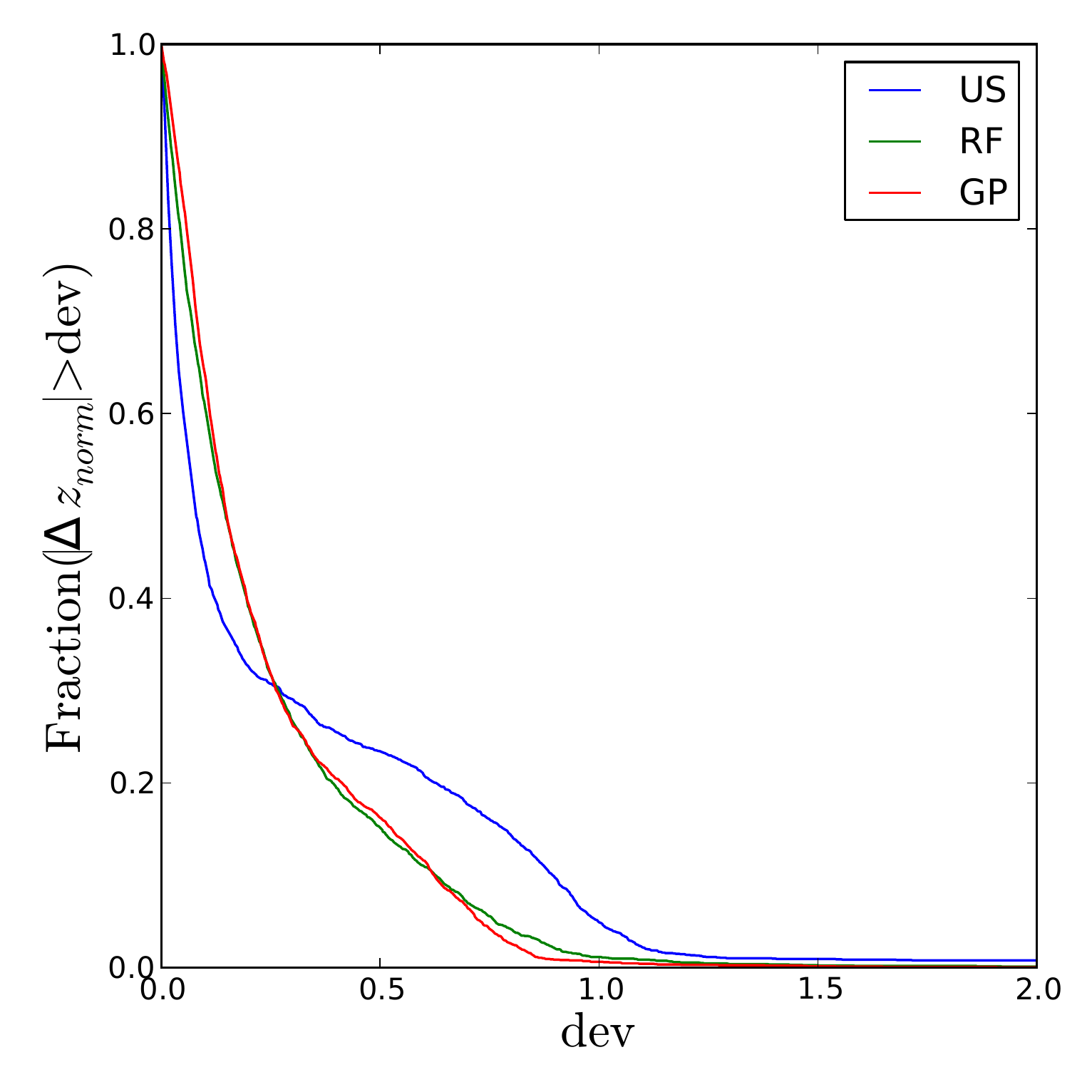}
\caption{The fraction of objects deviating with more than $|\Delta z_{norm}|>dev$ is plotted. 
Here, we only take a point prediction for our model $\argmax_z p_{spec}(z)$.
While the random forest and the Gaussian process show a very similar behaviour, the proposed algorithm can predict the redshift for 60\% of the objects significantly more precise but is heavily influenced from redshifts deviating more than 0.3.\label{fig:fractions}}
\end{figure}

The $MAD$ criterion is less sensitive to large deviations.
As a 
consequence, the $MAD$ measures rather the width of the central distribution than the width of the full
distribution. As seen in Fig.\,\ref{fig:allResults}, the predictions obtained from our model are much
more precise than the ones by the RF or GP. This becomes even clearer if we consider the fraction of objects
that deviate more than a certain value, cf. Fig.\,\ref{fig:fractions}. For the vast majority of the objects 
($\approx 70\%$) the deviation from the true value is considerably lower than for the RF and GP predictions. 

However, given that we are dealing with a problem where every set of magnitudes (or colours)
admits multiple solutions, the use of the $RMSE$ and $MAD$ is not appropriate as they focus
on comparing a point prediction to a single target solution\footnote{Which is fine for regression problems where an input is associated with a single target.} $z$.
Hence, $\frac{1}{N}logL_{\mathrm{True}}$ 
is better at quantifying performance as
it takes into account the fact that a model can predict multiple solutions $z$.
We therefore see that the proposed model displays a considerable higher likelihood than the RF and the GP.

\begin{table}
\caption{Summary of performance criteria.\label{tab:results}}
\centering
\begin{tabular}{lrrrr}
& Our approach & RF  & RF & GP \tabularnewline
&      & ($\sigma=0.30$) & ($\sigma=0.76$) & \tabularnewline
\hline
$RMSE$ & 0.476 & 0.344 & 0.344 & \textbf{0.326}\tabularnewline 
$MAD$ & \textbf{0.078} & 0.123 & 0.123 & 0.111 \tabularnewline
$\frac{1}{N}$logL$_{\mathrm{True}}$ & \textbf{-0.514} & -2.979 & -1.138 & -77.875\tabularnewline
\end{tabular}
\end{table}

\section{Discussion and Conclusion}

The presented approach can be developed further in two aspects: the methodological one and the astronomical one. 
Currently, as aforementioned, the coordinates $\bd{\theta}$ are discretised in order to prevent overfitting. It would be desirable to provide some control mechanism in order to constrain the coordinates $\bd{\theta}$ in regions of the lower dimensional space populated by data.
From an astronomical point of view, there is more work to be done. So far we have just demonstrated
the concept on a small dataset where the magnitudes were extracted with the provided filter curves and (known and noise-free) zero points were added. We chose this setting, as we wanted to have full control on the model and not to be distracted by erroneous and noisy calibrations. It is important to notice that a purely data-driven approach can deal with this quite naturally, while the presented algorithm depends heavily on the correctness of these calibrations. On the other hand, it is of course also possible to include a given uncertainty of the zero points into the model and also this can be cross-validated on a hold-out set. In summary, a much more detailed understanding of how photometric measurements relate to the spectra is 
required.

An advantage of our model is, that we can include uncertainty of the photometric measurements. This includes also \emph{missing} values which are a common struggle in astronomy
due to the different coverage and depth of the surveys. 
An interesting prospect is to extend the model towards the infrared. At the moment,
our coverage above $1\mu m$ is very shallow and thus it would be desirable to retrieve near-infrared to 
mid-infrared spectra of low-redshifted quasars (as otherwise the rest-frame would be in the optical again).
This would allow us to include also infrared data as then the coverage of the coordinates would reach into the 
near infrared. It is important to notice that it does not matter whether the infrared spectra are the same 
objects as the optical ones, it only has to be guaranteed that there is considerable overlap with the 
coordinates as they are now.  

Another issue is that standard regression models do not have control on the prior of $z$ and thus are implicitly biased by how $z$ is distributed in a given dataset.
Hence, this bias automatically propagated from training to future predictions as well. 
While this might be of advantage in some cases, it is a generally an unwanted side-effect of the training procedure. In contrast, the prior on $z$ is easily controlled in our model.

In conclusion,  we have presented an alternative way of modelling photometric redshifts. 
A simple model is formulated based on first principles and is given a probabilistic interpretation
which imparts a multimodal posterior distribution for redshifts. 
Numerical comparisons to regression models commonly used in this task, support our line of work.
We point out that by design, in their standard formulation, the RF and GP cannot fully address the problem of redshift estimation as they cannot by design predict 
multiple distinct solutions. One possibility is perhaps to extend these techniques by taking inspiration
from the mixture density networks \cite{MDN} or mixture of experts \cite{Jordan1994}.


\section*{Acknowledgment}
The authors gratefully acknowledge the support of the Klaus Tschira Foundation.
The authors would also like to thank Pavlos Protopapas for his input
in formulating the physical model and also acknowledge SDSS 
for their valuable service in making data access possible.
Funding for the Sloan Digital Sky Survey IV has been provided by
the Alfred P. Sloan Foundation, the U.S. Department of Energy Office of
Science, and the Participating Institutions. SDSS-IV acknowledges
support and resources from the Center for High-Performance Computing at
the University of Utah. The SDSS web site is www.sdss.org.



\bibliographystyle{unsrt}
\bibliography{mybiblio}
%

\end{document}